\documentclass[preprint,review,12pt]{elsarticle}

\usepackage{graphicx}
\usepackage{amssymb}
\usepackage{lineno}
\usepackage[super]{nth}
\usepackage[dvipsnames]{xcolor}

\journal{JASTP}

\begin{document}

\begin{frontmatter}


\title{On the Low-Latitude Ionospheric Irregularities under Geomagnetically Active and Quiet Conditions using NavIC observables: A Spectral Analysis Approach}



\author[a,b]{Sumanjit Chakraborty\corref{c-d54cc1eb1ca4}}
\ead{sumanjit11@gmail.com}\cortext[c-d54cc1eb1ca4]{Corresponding author.}
\author[a]{Abhirup Datta}
\ead{abhirup.datta@iiti.ac.in}
\address[a]{Department of Astronomy, Astrophysics and Space Engineering, Indian Institute of Technology Indore, Simrol, Indore 453552, Madhya Pradesh, India}
\address[b]{Indian Institute of Geomagnetism, Navi Mumbai, India}

\begin{abstract}

Ionospheric irregularities and associated scintillations under geomagnetically active/quiet conditions have detrimental effects on the reliability and performance of space- and ground-based navigation satellite systems, especially over the low-latitude region. The current work investigates the low-latitude ionospheric irregularities using the phase screen theory and the corresponding temporal Power Spectral Density (PSD) analysis to present an estimate of the outer irregularity scale sizes over these locations. The study uses simultaneous L5 signal C/N$_o$ observations of NavIC (a set of GEO and GSO navigation satellite systems) near the northern crest of EIA (Indore: 22.52$^\circ$N, 75.92$^\circ$E, dip: 32.23$^\circ$N) and in between the crest and the dip equator (Hyderabad: 17.42$^\circ$N, 78.55$^\circ$E, dip: 21.69$^\circ$N). The study period (2017-2018) covers disturbed and quiet-time conditions in the declining phase of the solar cycle 24. The PSD analysis brings forward the presence of irregularities, of the order of a few hundred meters during weak-to-moderate and quiet-time conditions and up to a few km during the strong event, over both locations. The ROTI values validate the presence of such structures in the Indian region. Furthermore, only for the strong event, a time delay of scintillation occurrence over Indore, with values of 36 minutes and 50 minutes for NavIC satellites (PRNs) 5 and 6, respectively, from scintillation occurrence at Hyderabad is observed, suggesting a poleward evolution of irregularity structures. Further observations show a westward propagation of these structures on this day. This study brings forward the advantage of utilizing continuous data from the GEO and GSO satellite systems in understanding the evolution and propagation of the ionospheric irregularities over the low-latitude region. 

\end{abstract}

\begin{keyword}

Ionospheric Scintillation \sep Ionospheric Irregularities \sep NavIC \sep Power Spectral Density \sep ROTI


\end{keyword}

\end{frontmatter}


\section{Introduction}

The phenomenon of scintillation can be thought of as analog to the stars that twinkle in the night sky as a result of variations in the density of the atmosphere due to turbulence. The phenomenon of plasma instability (post-sunset) in the equatorial ionosphere generates irregularities and large-scale depletions in the electron density referred to as the Equatorial Plasma Bubbles (EPBs) (\cite{sc:18} and references therein). Radio waves propagating through these irregularities experience diffraction and scattering, which cause random fluctuations in the VHF and L-band signal amplitude, phase, the direction of propagation, and polarization referred to as scintillations \citep{sc:3,sc:4,sc:8}. It is well known that the irregularities that produce scintillations are mainly found in the F layer of the ionosphere ($\sim$ 300-350 km), where the plasma density has the maximum value. Rayleigh–Taylor (R-T) instabilities are considered as the primary mechanism generating EPBs with scale sizes from a few hundred meters to several kilometers \citep{sc:8}.
Activities of scintillations are generally observed during the period of maximum solar activity, occurring near the magnetic equator in the post-sunset-to-midnight sector \citep{sc:7,sc:6}. It is well known that ionospheric scintillations are demonstrations of space weather effects, affecting the performance of space-based navigation and communication systems that rely on transionospheric radio-wave propagation \citep{sc:4,sc:5}. 

For characterizing radio wave scintillation that travels in the ionosphere, numerous phase screen theories have been developed, as early as the 1950s following works done by \citep{sc:1,sc:2}. Generally, the scintillation of radio waves is calculated using the theory of wave propagation in random media to obtain the parameters of the exiting wave from the ionosphere. It is performed by solving the Fresnel diffraction theory problem involving the propagation of these waves traveling between the ionosphere and the Earth. Thus, these calculations replace the ionosphere with an equivalent phase screen that is random, where the irregularities in the ionosphere are considered as phase objects \citep{sc:10}. Simultaneously, the ground pattern produced by radio waves propagating through this screen is derived from the diffraction theory \citep{sc:11}. In general, the one- and two-dimensional phase screens are analytic representations of the intensity of the spectral density function in terms of the various phase structure function combinations \citep{sc:9}. Phase screen models have been extensively used to simulate the distortion of wideband waveforms for communications \citep{sc:26}, fading of GNSS satellite signals, and scintillation of satellite signals used in radio occultation experiments, to name a few. These phase screen models' inputs are the in-situ electron density measurements, a radio receiver's time series phase measurements, and a turbulent ionospheric medium-based stochastic model. 

Several researchers (\cite{sc:3,sc:4,sc:5,sc:29,sc:30} and references therein) have addressed in-situ electron density fluctuations measured by the satellites to predict the propagation effects. However, the disadvantage lies in the fact that measured values sample the electron density irregularities and fluctuations at the satellite's orbital altitude. The transionospheric propagation effects can be attributed to the integrated development of the density variation along the signal ray path at all altitudes. If any, the relation between density fluctuations sampled at one altitude to another is still not evident in the literature. Therefore, if the density irregularities associated with an EPB have not risen to the satellite's altitude that measures the in-situ density values, detection of fluctuations will be absent despite the EPB causing radio scintillations as a result of the irregularities present at lower altitudes. There is an advantage of probing the ionosphere at all altitudes, from the ground-based observations of signal phase transmitted by a satellite at the topside ionosphere or beyond, over in-situ electron density observations. Thus, it becomes essential to understand and quantify the contribution from phase scintillations caused by diffraction, in addition to interpreting the signal phase time series and the corresponding limitations of judiciously representing the structure of irregularity at these altitudes \citep{sc:25,sc:4,sc:21,sc:20,sc:23,sc:24}.

The presence of the northern crest of the Equatorial Ionization Anomaly (EIA) and the geomagnetic equator that touches the southern tip of the Indian peninsular region, accompanied by sharp latitudinal gradients of ionization, make the Indian longitude sector a highly geosensitive region of investigation for ionospheric research during geomagnetically disturbed periods when the low-latitude ionization is significantly affected as a result of solar eruptions like the Coronal Mass Ejections (CMEs) \citep{sc:19,sc:55,sc:56,sc:57,sc:37}, the coronal hole-high speed solar wind streams associated co-rotating interaction regions \citep{sc:12}. Although previous studies show the development of power law phase screen theories and models for scintillation using the GPS/GNSS, the nature of ionospheric irregularities using the Navigation with Indian Constellation (NavIC) in the low-latitude region of India, during geomagnetically active as well as quiet-time conditions, has not been studied in the literature. The novelty of this work lies in the simultaneous investigation of the evolution and propagation of low-latitude ionospheric irregularities, during the strong event, over a location (Indore) near the northern crest of the EIA and a location (Hyderabad) in between the crest and the magnetic equator in Indian sector, utilizing the carrier-to-noise C/N$_o$ observations from a set of geostationary (GEO) and geosynchronous (GSO) satellite system of NavIC.  

The manuscript is presented as follows: Section 2 briefly describes the dataset used. Section 3 shows the results and the PSD analysis of the three cases of strong, moderate, and weak event days. Section 4 presents a discussion of these results, while Section 5 presents the summary.

\section{Data}

The NavIC is a regional navigation satellite system developed by the Indian Space Research Organisation (ISRO). The space segment of NavIC consists of a combination of seven GEO and GSO satellites. It provides continuous (spatial and temporal) monitoring of the upper atmosphere over the low-latitude and equatorial regions in the Indian longitude sector. It is conceived to provide accurate positioning information to all users in and around the 1500 km radius of the boundaries of the country. The suitability of using NavIC for studying the upper atmosphere, over the Indian subcontinent in total and near the EIA (Indore) in particular, has been well established by \citep{sc:35,sc:13,sc:28,sc:36,sc:34,sc:32,sc:33,sc:31}.     

\begin{figure}[ht]
\noindent\includegraphics[width=5in,height=5in]{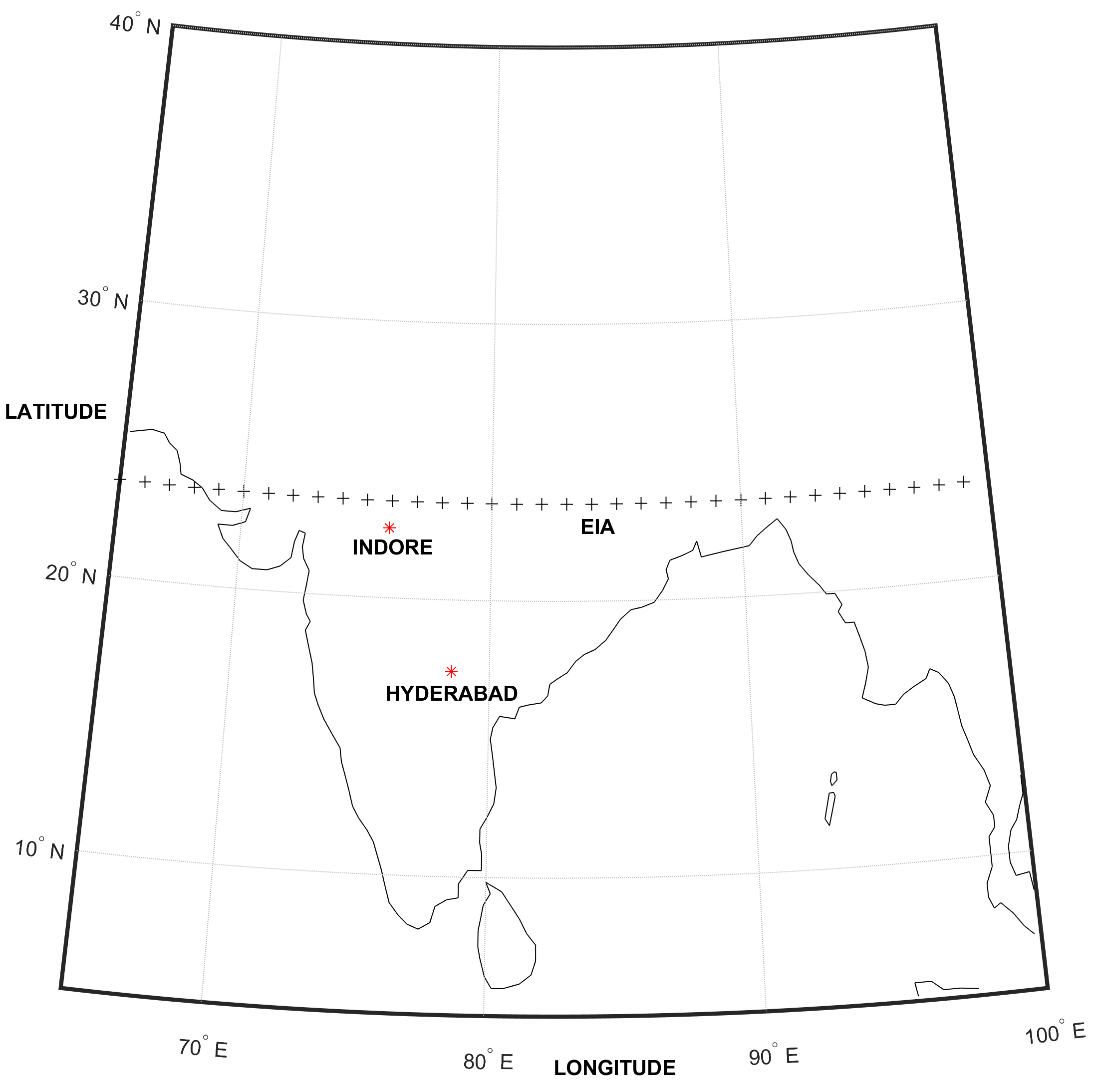}
\caption{The map of India showing the locations of the two stations and the EIA crest.} 
\label{fig1}
\end{figure}

A NavIC receiver is operational at the Department of Astronomy, Astrophysics and Space Engineering of the Indian Institute of Technology Indore (22.52$^\circ$N, 75.92$^\circ$E geographic; magnetic dip 32.23$^\circ$N) since April 2017. It is capable of receiving L5 (1176.45 MHz) and S1 (2492.028 MHz) signals along with GPS L1 (1575.42 MHz) signals. The output of the receiver includes the carrier-to-noise C/N$_o$ (dB-Hz), azimuth (deg), elevation (deg), pseudo-range (in m), and carrier cycles (cycles). The receiver is provided by the Space Applications Centre, ISRO. Additionally, data from another NavIC receiver operational at the Department of Electronics and Communication Engineering (ECE), Osmania University (OU), Hyderabad (17.42$^\circ$N, 78.55$^\circ$E geographic; magnetic dip 21.69$^\circ$N), has been used. A similar procedure as shown by \citep{sc:28} has been followed here for the analysis of these NavIC data. Figure \ref{fig1} shows the map that contains the locations of observations and the northern crest of the EIA that passes over the Indian sector.

\clearpage

\section{Results}

In this section, the strong storm day of September 08, 2017, the moderate storm day of September 16, 2017, and the weak/minor storm day of August 17, 2018, are presented as sample cases from the 27 days of observations of scintillations during one year, to study the C/N$_o$ variation as observed by the NavIC satellites. The geomagnetic storm intensity in this study is classified following \citep{sc:54}. The analyses of the corresponding PSDs are then performed and calculations of the spectral slope and estimation of the outer scale sizes of irregularities, over Indore and Hyderabad, are presented. It is to be noted that as NavIC satellites are at geostationary altitudes, the C/N$_o$ data can be continuously observed throughout the day.

\subsection{September 08, 2017: a strong storm day}

\begin{figure}[ht]
\noindent\includegraphics[width=5in,height=5in]{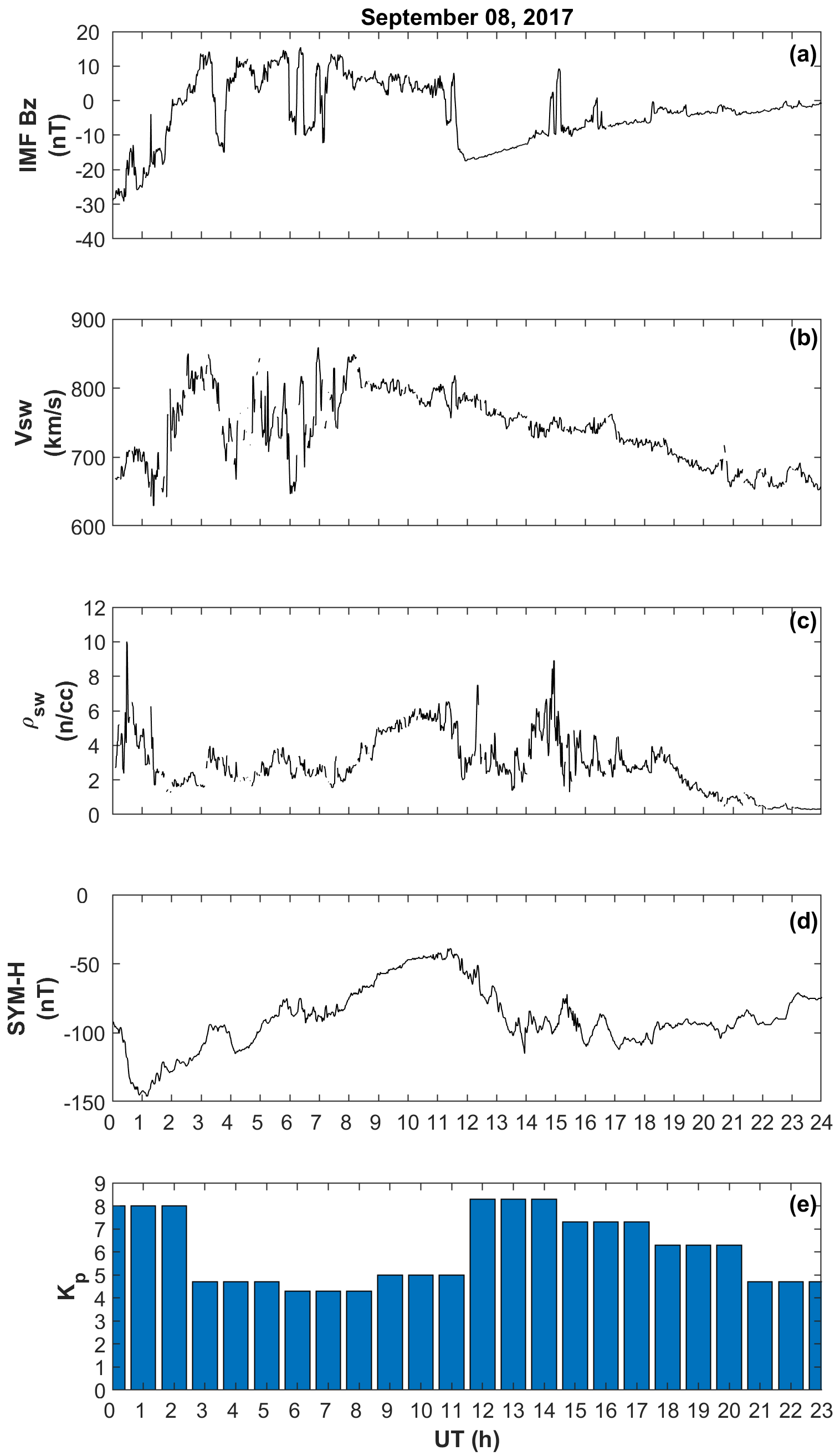}
\caption{Variations of (a) IMF B$_z$ (nT), (b) V$_{sw}$ (km/s), (c) $\rho_{sw}$ (n/cc), (d) SYM-H (nT) and (e) K$_p$ during September 08, 2017.} 
\label{fig2}
\end{figure}

Due to a CME that arrived on September 06, 2017, a G4 (K$_p$= 8, severe) level geomagnetic storm was observed at 23:50 UT on September 07, 2017, at 01:51 UT and 13:04 UT on September 08, 2017, according to the NOAA (https://www.swpc.noaa.gov/). Figure \ref{fig2} shows the interplanetary and geomagnetic conditions on September 08, 2017. The geomagnetic storm had been strong and had a double-peak, evident from Figure \ref{fig2} (d) showing drops in the SYM-H (nT) values to -146 nT at 01:08 UT and -112 nT at 17:07 UT on the day. The corresponding IMF B$_z$ (nT) in Figure \ref{fig2} (a) showed values of -20.69 nT and -6.12 nT respectively, suggesting strong southward IMF conditions favorable for an intense geomagnetic storm. The solar wind flow speed, (V$_{sw}$, km/s), the solar wind density ($\rho_{sw}$, n/cc), and the $K_p$ values in Figure panels \ref{fig2} (b), (c), and (e) respectively showed values of 694.3 km/s, 2.91 n/cc and 8 at 01:08 UT and 718 km/s, 3.14 and 7+ at 17:07 UT on September 08, 2017. 
\clearpage

\clearpage

\begin{figure}[ht]
\noindent\includegraphics[width=5in,height=5in]{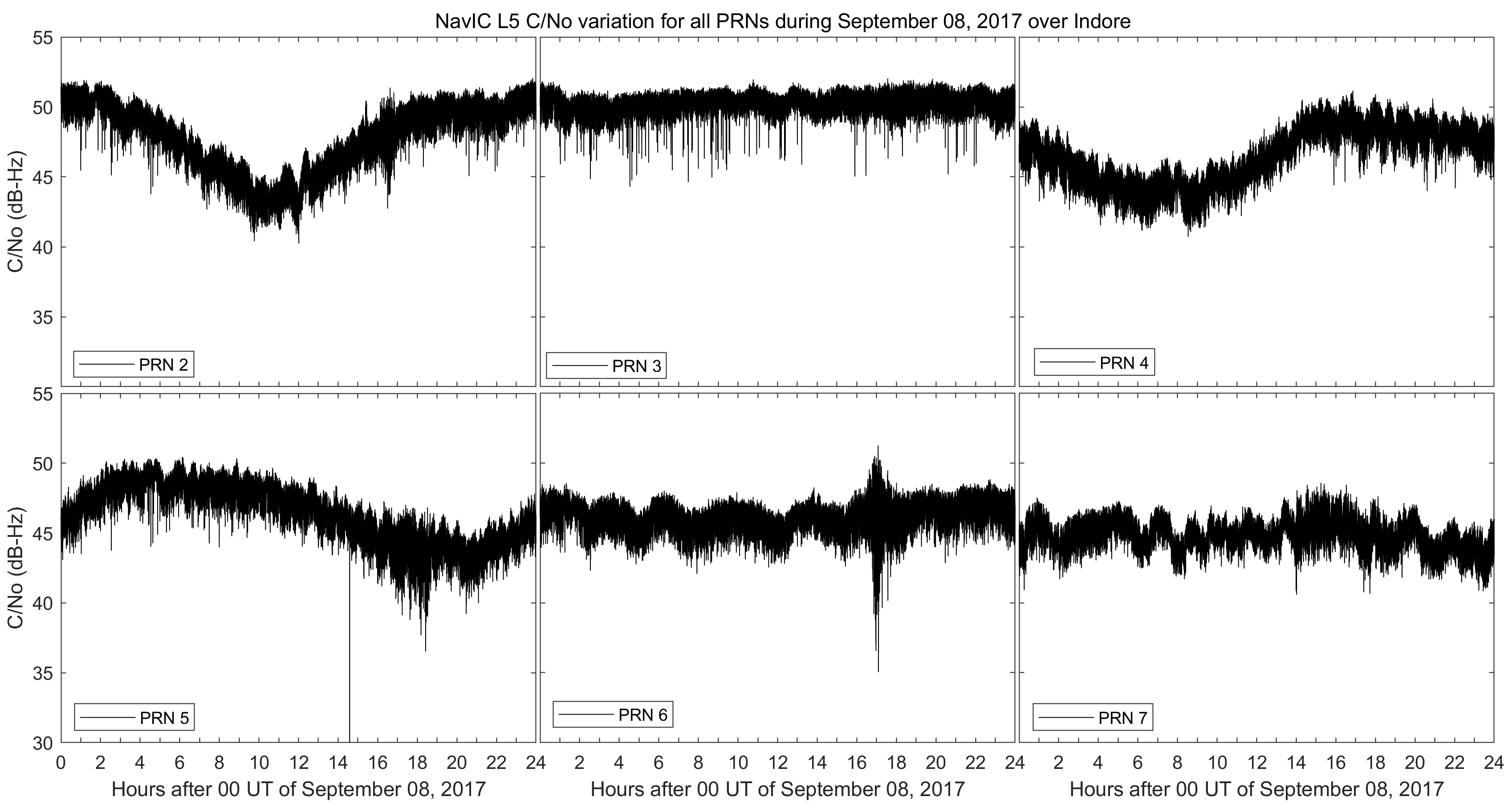}
\caption{The C/N$_o$ (dB-Hz) variation during the disturbed day of September 08, 2017, as observed by the L5 signal of NavIC satellite PRNs 2-7 over Indore. It is to be noted that the LT = UT + 05:04 h.} 
\label{fig3}
\end{figure}

Figure \ref{fig3} shows the C/N$_o$ variation (dB-Hz) over Indore located near the northern crest of the EIA that passes over the highly geosensitive Indian longitude sector, for the entire day of September 08, 2017, as observed by a set of geostationary (PRNs 3, 6 and 7) and geosynchronous (PRNs 2, 4 and 5) satellites of NavIC. The observations are taken by utilizing the L5 (1176.45 MHz) signal of NavIC. Upon closely observing panels showing PRNs 5 and 6 of Figure \ref{fig3}, it is visible that there had been strong fluctuations in the C/N$_o$ with values dropping from about 47 dB-Hz to 36 dB-Hz around 18-19 UT (23:04-00:04(+1) LT) for PRN 5 and from about 51 dB-Hz to 35 dB-Hz around 17-18 UT (22:04-23:04 LT) for PRN 6, which are in between the local pre-midnight and post-midnight sector. It is to be noted that the high value observed in the 14-15 UT bin for PRN5 has not been considered as it is getting manifested as a result of the C/N$_o$ drop to zero observed by this PRN. 

\begin{figure}[ht]
\centering
\noindent\includegraphics[width=4.5in,height=4.5in]{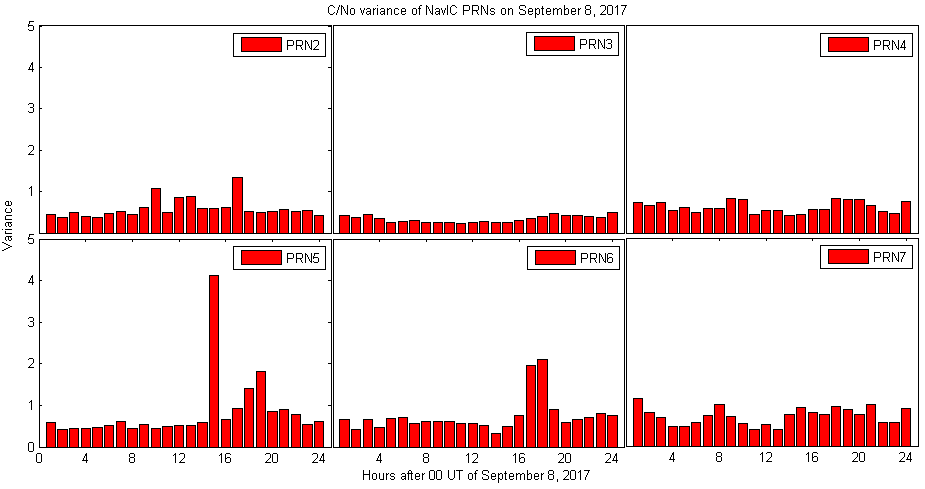}
\caption{The hourly binned variance plots of C/N$_o$ for all PRNs of NavIC on September 08, 2017, as observed from Indore. It is to be noted that the LT = UT + 05:04 h.} 
\label{fig4}
\end{figure}

To verify whether the observed C/N$_o$ drops in Figure \ref{fig3} are due to the scintillations that had occurred during this period, Figure \ref{fig4} shows the hourly binned variance plots of all the PRNs of NavIC during the entire day of September 08, 2017. In bin 18-19 UT for PRN 5 and bin 17-18 UT for PRN 6, the variance over the average values is observed to be significant.
\clearpage

The theory of temporal power spectrum has been introduced by several researchers \citep{sc:14,sc:15,sc:16,sc:17} as follows: 
\begin{linenomath*}
\begin{equation}
S = \frac{T}{(f_0^2 + f^2)^\frac{p}{2}}
\label{sca}
\end{equation}
\end{linenomath*}
where f$_0$ is the outer scale frequency, T is the spectral strength and p is the spectral slope. Equation \ref{sca} can be simplified to the following (when $f >> f_0$):
\begin{linenomath*}
\begin{equation}
S = T f^{-p}
\label{scb}
\end{equation}
\end{linenomath*}
Furthermore, the spectrum of electron density fluctuations ($\delta N$) can be modeled as a power law, with an outer scale, given by \citep{sc:15,sc:16,sc:22}:
\begin{linenomath*}
\begin{equation}
S_{\delta N}(q) =  \frac{C_s}{(q_0^2+q^2)^{(m+\frac{1}{2})}}
\label{scc}
\end{equation}
\end{linenomath*}
where, $C_s$ is the strength of the turbulence proportional to T \citep{sc:15,sc:16,sc:22}, m is the irregularity spectral index (p = 2m+1) and $q_0$ is the wave number of the outer scale and is related to the outer scale turbulence (L) as:
\begin{linenomath*}
\begin{equation}
L = \frac{2\pi}{q_0}
\label{scd}
\end{equation}
\end{linenomath*}
Therefore, for $q>>q_0$, the spectrum in equation \ref{scc} modifies to:

\begin{linenomath*}
\begin{equation}
S_{\delta N}(q) = C_sq^{-(2m+1)} = C_sq^{-p}
\label{sce}
\end{equation}
\end{linenomath*}

\begin{figure}[ht]
\centering
\noindent\includegraphics[width=2.5in,height=2.5in]{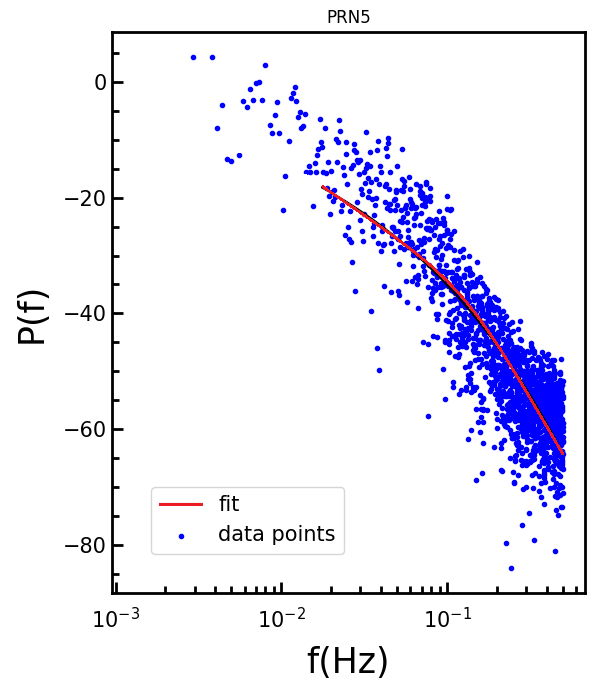}
\noindent\includegraphics[width=2.5in,height=2.5in]{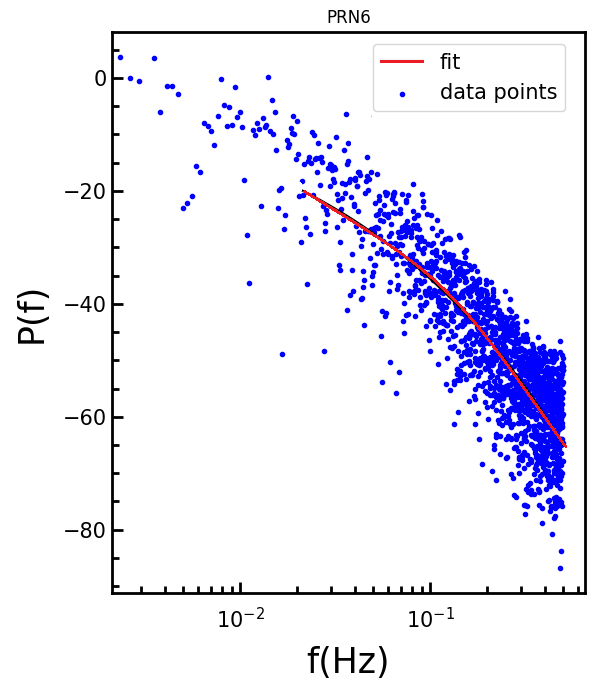}
\caption{The PSD variations with the least square fit (red solid line) corresponding to the bins of intense C/N$_o$ variation of Figure \ref{fig4}.} 
\label{fig5}
\end{figure}

Figure \ref{fig5} shows the PSD variation from PRN5 and PRN6. The significant time bins (18-19 UT and 17-18 UT for PRNs 5 and 6 respectively) are used for obtaining the PSD variations in this figure. The absolute values of p for the PRNs 5 and 6 are 3.690$\pm$0.009 and 3.596$\pm$0.008 respectively. These values are obtained from the covariance matrix of the least squares fit performed to the PSD in the logarithmic domain by using equation \ref{scb}. These values are then verified with the approximate estimation \citep{sc:15,sc:16} of $p = \frac{[P(0.05)-P(0.5)]}{10}$ from Figure \ref{fig5}. Further, utilizing equation \ref{sce} followed by equation \ref{scd}, in addition to these spectral slope values from the PSD analysis from equation \ref{scb}, the outer scale of the irregularities is obtained to be 5.12$\pm0.011$ km.    

Taking a similar approach for the storm day of September 08, 2017, the following Figure \ref{fig6} shows the NavIC measured C/N$_o$ variation (dB-Hz) over Hyderabad. Upon closely observing panels showing PRNs 5 and 6, it is visible that there had been strong fluctuations in the C/N$_o$ with values dropping from about 50 dB-Hz to 36 dB-Hz around 17-18 UT (22:14-23:14 LT) for PRN 5 and from about 50 dB-Hz to 37 dB-Hz around 16-17 UT (21:14-22:14 LT), both during the local pre-midnight sector.
\begin{figure}[ht]
\centering
\noindent\includegraphics[width=4.5in,height=4.5in]{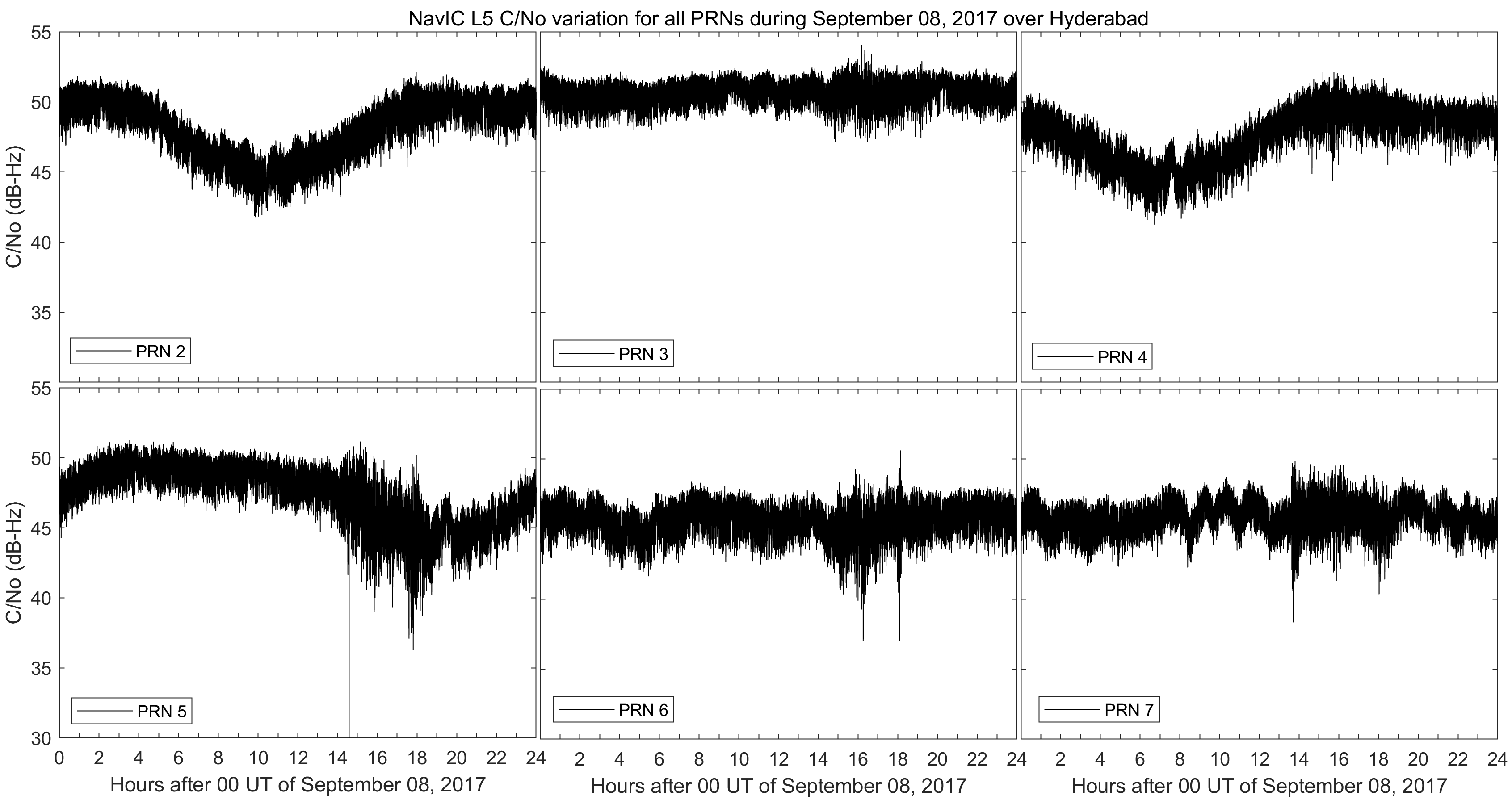}
\caption{The C/N$_o$ (dB-Hz) variation during the disturbed day of September 08, 2017, as observed by the L5 signal of NavIC satellite PRNs 2-7 over Hyderabad. It is to be noted that the LT = UT + 05:14 h.}
\label{fig6}
\end{figure} 

\begin{figure}[ht]
\centering
\noindent\includegraphics[width=4.5in,height=4.5in]{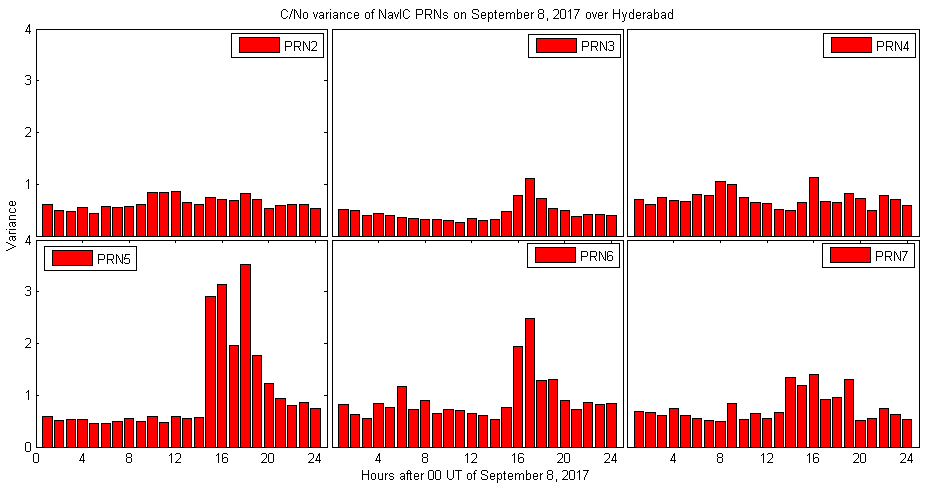}
\caption{The hourly binned variance plots of C/N$_o$ for all PRNs of NavIC on September 08, 2017, over Hyderabad. It is to be noted that the LT = UT + 05:14 h.} 
\label{fig7}
\end{figure}

\begin{figure}[ht]
\centering
\noindent\includegraphics[width=2.5in,height=2.5in]{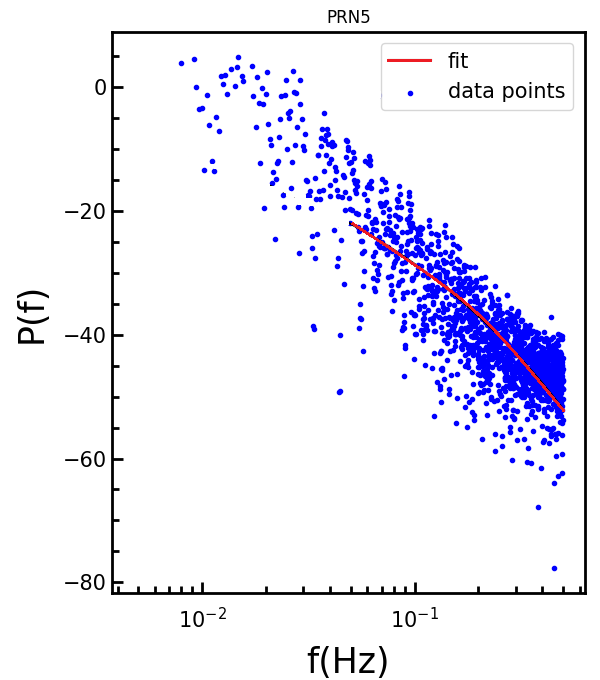}
\noindent\includegraphics[width=2.5in,height=2.5in]{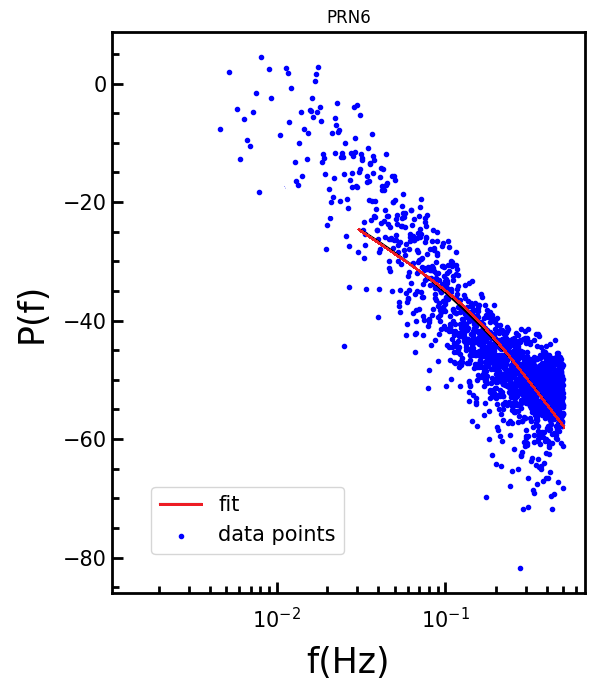}
\caption{The PSD variations with the least square fit (red solid line) correspond to the time bins of intense C/N$_o$ variation of Figure \ref{fig7}.} 
\label{fig8}
\end{figure}

Similarly, for verification of the observed C/N$_o$ drops due to scintillation, Figure \ref{fig7} shows the hourly binned variance plot for all the PRNs of NavIC during the entire day of September 08, 2017. The variance over the average values is significant in the hourly bin of 17-18 UT for PRN 5 and 16-17 UT for PRN 6.
Figure \ref{fig8} shows the PSD variation from the PRNs 5 and 6. The corresponding values of p, using the above equations, for PRNs 5 and 6 are 3.072$\pm$0.011 and 3.008$\pm$0.010 respectively. The corresponding outer scale size of irregularities obtained is 5.15$\pm0.001$ km. 
\clearpage

\subsection{September 16, 2017: a moderate storm day}

\begin{figure}[ht]
\includegraphics[width=5in,height=5in]{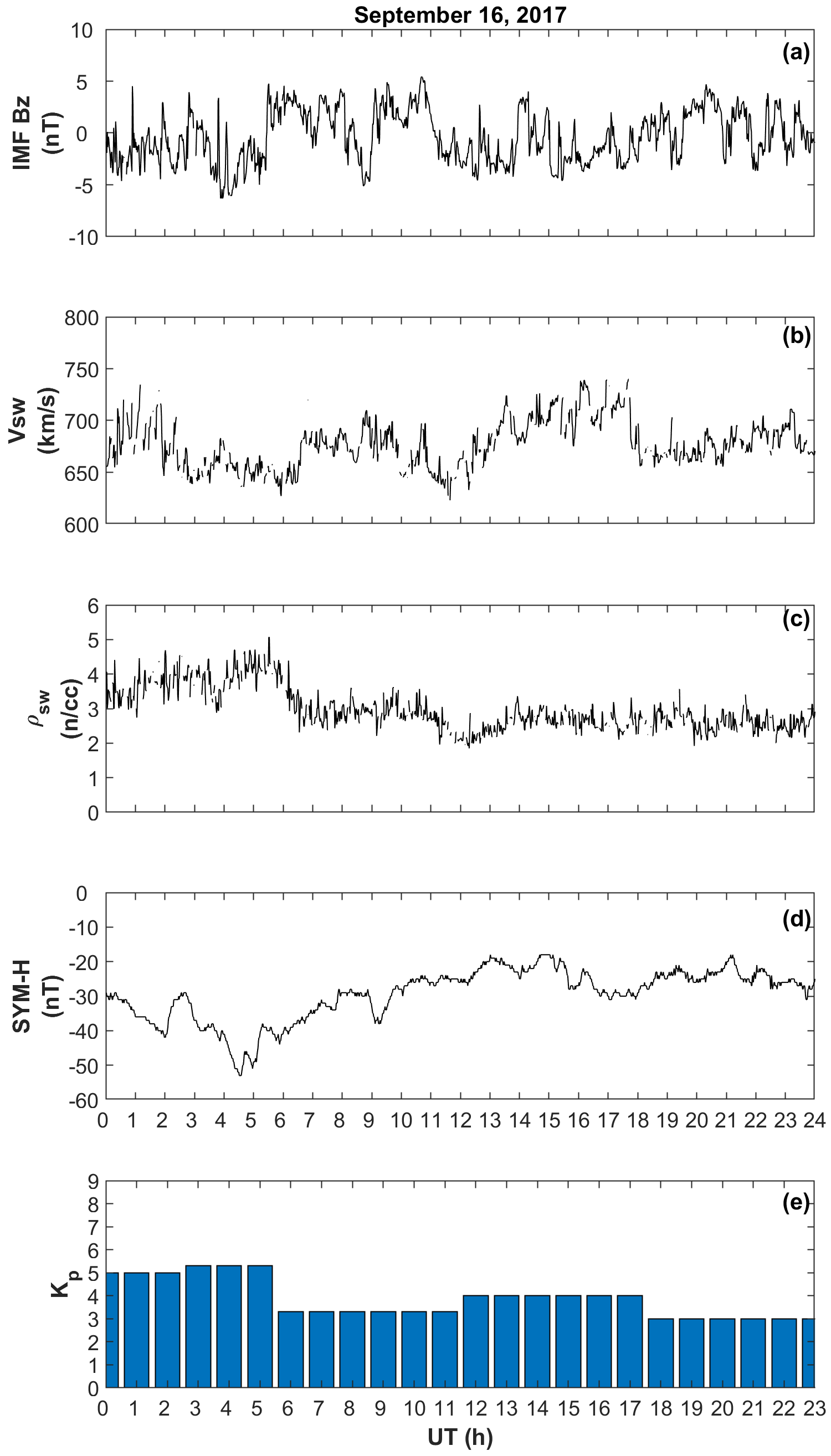}
\caption{Variations of (a) IMF B$_z$ (nT), (b) V$_{sw}$ (km/s), (c) $\rho_{sw}$ (n/cc), (d) SYM-H (nT) and (e) K$_p$ during September 16, 2017.} 
\label{fig9}
\end{figure}

The high-speed solar wind blew past Earth at a velocity of around 650-850 km/s. This sparked a G2-class (NOAA Space Weather Scales) moderate geomagnetic storm on September 16, 2017. Figure \ref{fig9} shows the interplanetary and geomagnetic conditions on September 16, 2017. Corresponding to the SYM-H drop (Figure \ref{fig9} (d)), the IMF B$_z$, the V$_{sw}$, the $\rho_{sw}$, and the $K_p$ values in Figure panels \ref{fig9} (a), (b), (c), and (e) showed the values of -2.59 nT, 694.3 km/s, 3.56 n/cc and 5+ respectively on September 16, 2017.     

Following a similar approach as presented for the previous case, the following Figure \ref{fig10} shows the C/N$_o$ variation (dB-Hz) for the entire day of September 16, 2017, as observed by the L5 signal of NavIC. Drops in the C/N$_o$ have been observed at multiple time stamps throughout the day by all the PRNs.

However, to confirm whether the C/N$_o$ drops had been significant, Figure \ref{fig11} has been plotted that shows the hourly binned variance plots of all the PRNs of NavIC during the entire day of September 16, 2017. The hourly bin of 6-7 UT for all the PRNs shows the most significant rise and hence maximum variation among all the bins of the day.

Figure \ref{fig12} shows the PSD variation from all the satellites of NavIC. It is to be noted that only PRNs 3, 4, and 5 show a power law variation in the PSD and hence are taken into consideration. The values of p for PRNs 3, 4, and 5 are 2.873$\pm$, 2.949$\pm$0.009, and 2.829$\pm$0.008 respectively. The corresponding outer irregularity scale size is 0.507$\pm$0.001 km.

\clearpage

\begin{figure}[ht]
\centering
\noindent\includegraphics[width=4.5in,height=4.5in]{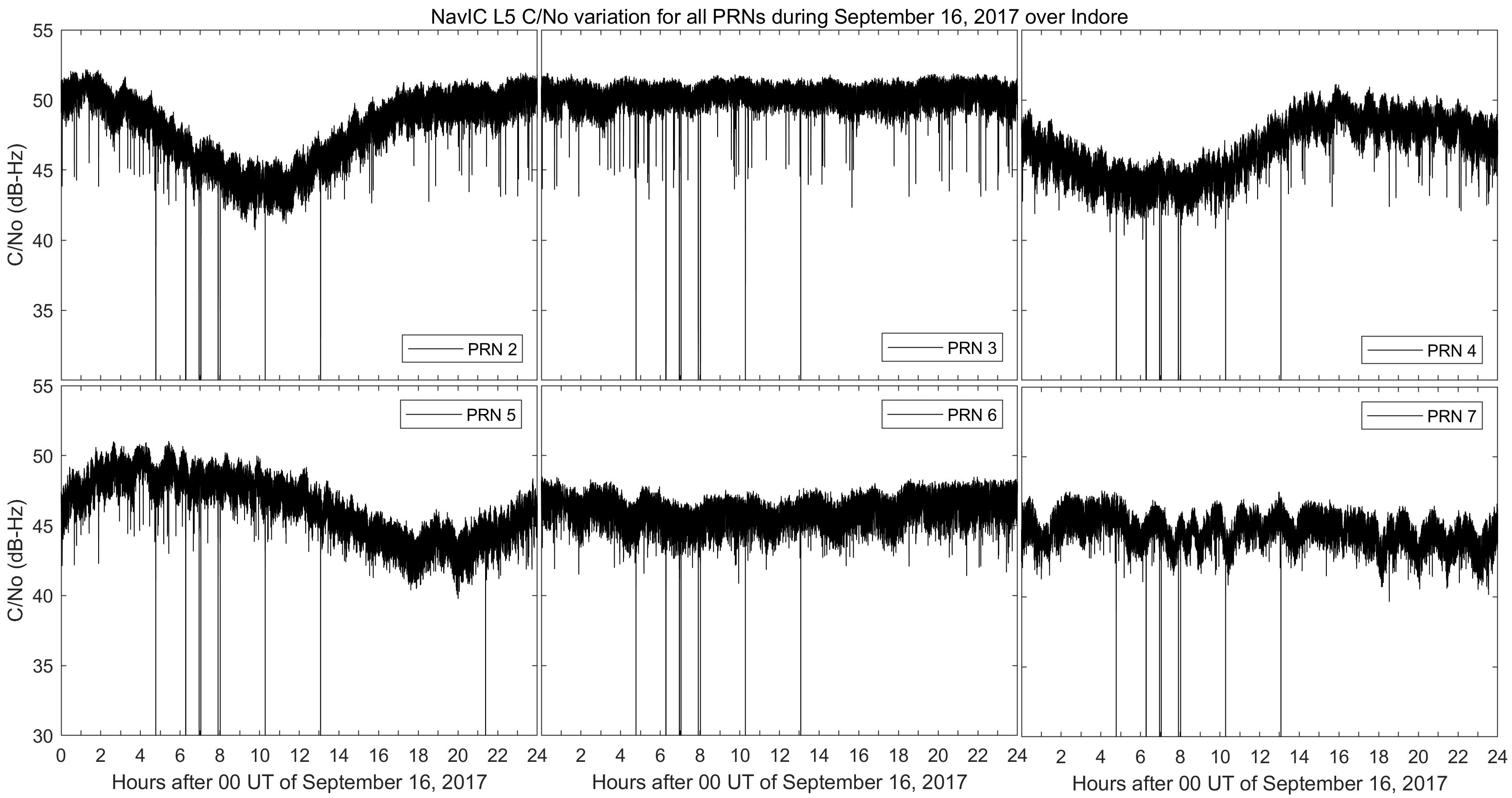}
\caption{The C/N$_o$ (dB-Hz) variation during the entire day of September 16, 2017, over Indore, as observed by the L5 signal of NavIC satellite PRNs 2-7.} 
\label{fig10}
\end{figure}

\begin{figure}[ht]
\centering
\noindent\includegraphics[width=4.5in,height=4.5in]{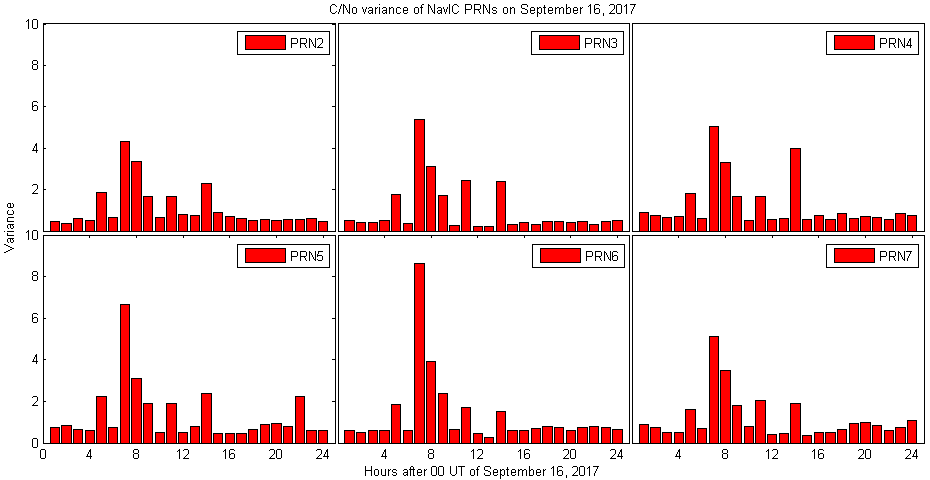}
\caption{The hourly binned variance plots of C/N$_o$ for all PRNs of NavIC on September 16, 2017, over Indore.} 
\label{fig11}
\end{figure}

\begin{figure}[ht]
\centering
\noindent\includegraphics[width=3in,height=2.2in]{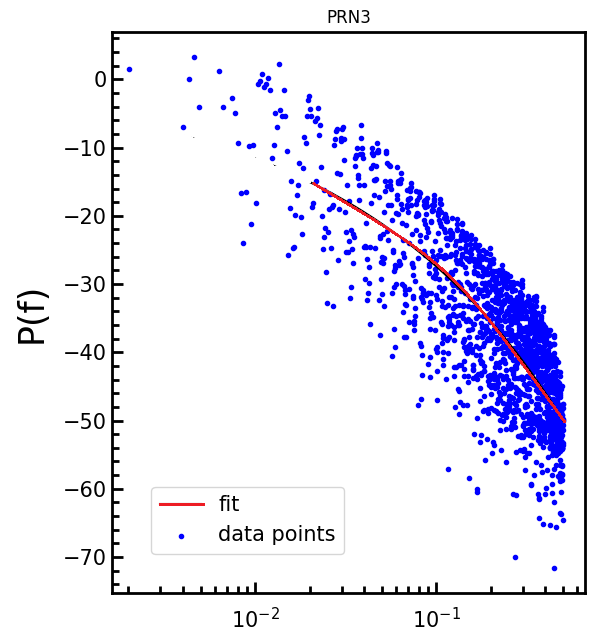}
\noindent\includegraphics[width=3in,height=2.2in]{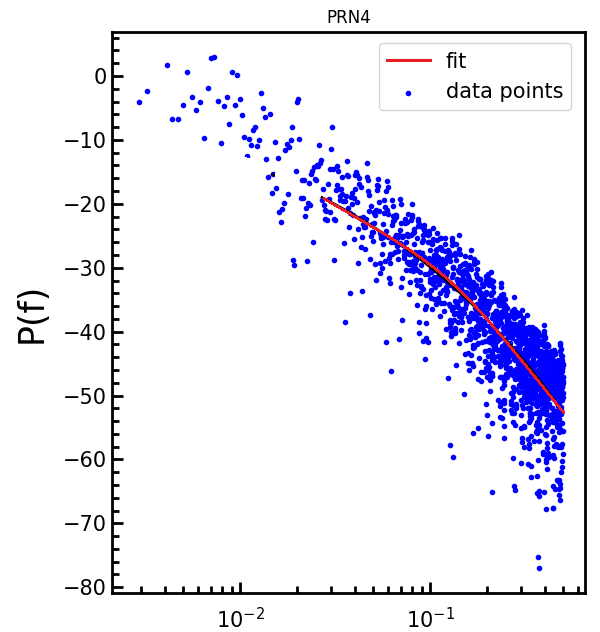}
\noindent\includegraphics[width=3in,height=2.2in]{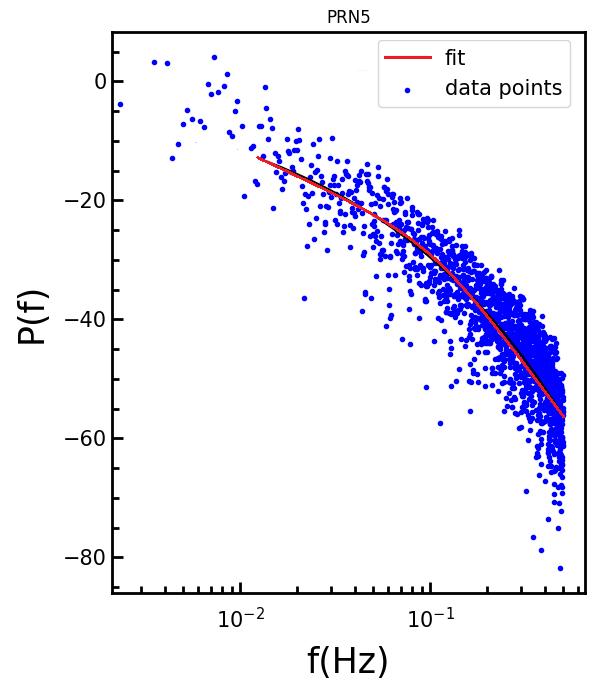}
\caption{The PSD variations with the least square fit (red solid line) correspond to the time bins of C/N$_o$ variation of Figure \ref{fig11}.} 
\label{fig12}
\end{figure}
\clearpage

\subsection{August 17, 2018: a weak/minor storm day}

\begin{figure}[ht]
\includegraphics[width=5in,height=5in]{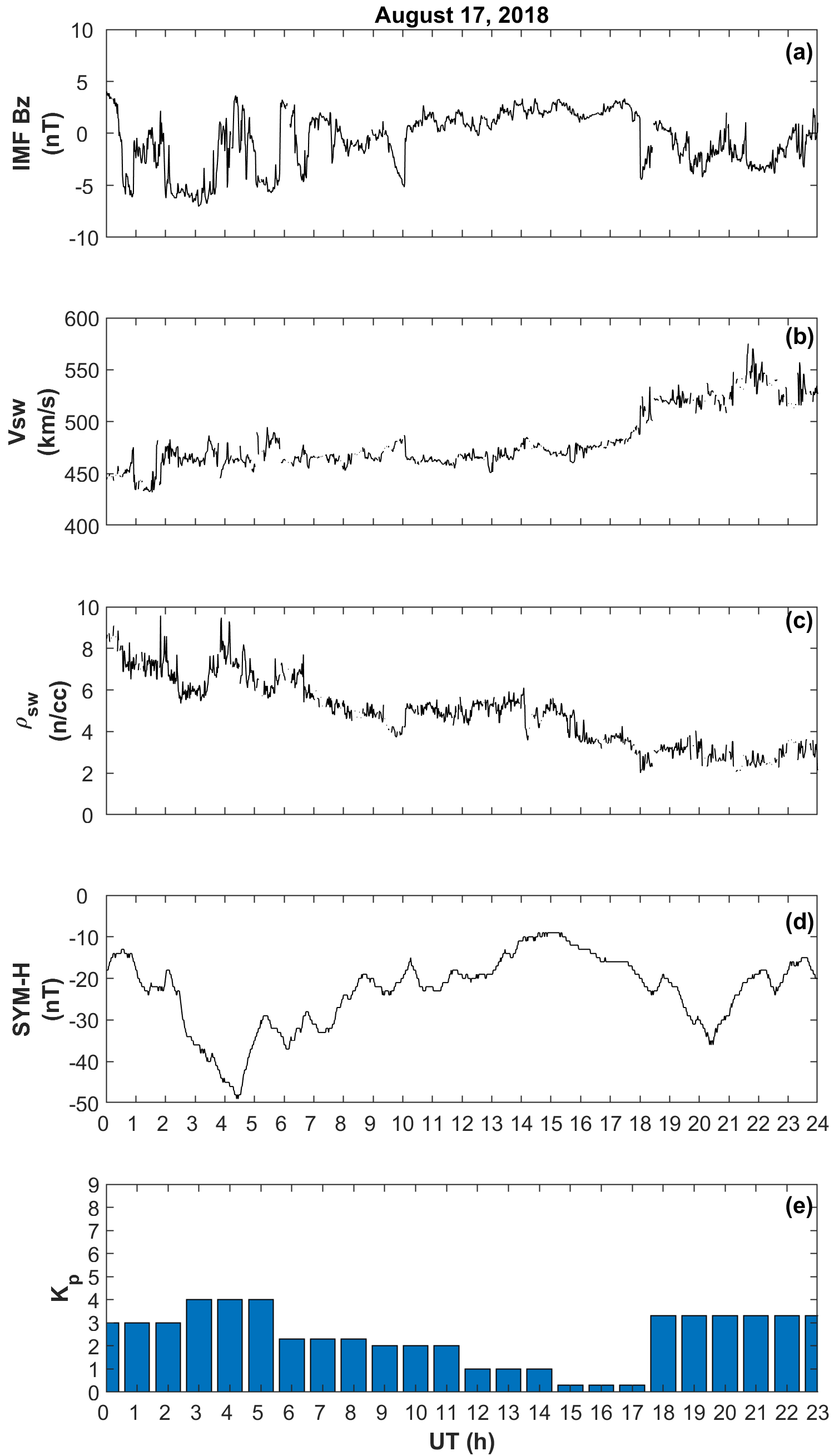}
\caption{Variations of (a) IMF B$_z$ (nT), (b) V$_{sw}$ (km/s), (c) $\rho_{sw}$ (n/cc), (d) SYM-H (nT) and (e) K$_p$ during August 17, 2018.} 
\label{fig13}
\end{figure}

A solar wind stream hit the geomagnetic field during the late hours of August 15, 2018, and blew throughout the next day. This sparked a G1-class (NOAA Space Weather Scales) weak/minor geomagnetic storm on August 17, 2018. Figure \ref{fig13} shows the interplanetary and geomagnetic conditions on August 17, 2018. The corresponding values of IMF B$_z$, the V$_{sw}$, the $\rho_{sw}$, and the $K_p$ (shown in Figure panels \ref{fig13} (a), (b), (c), and (e) respectively), for the SYM-H drops (Figure \ref{fig13} (d)) are 3.36 nT, 463.5 km/s, 7.50 n/cc, 4 and -2.78 nT, 525.4 km/s, 2.91 n/cc, 3+ respectively on August 17, 2018.

\begin{figure}[ht]
\noindent\includegraphics[width=5in,height=5in]{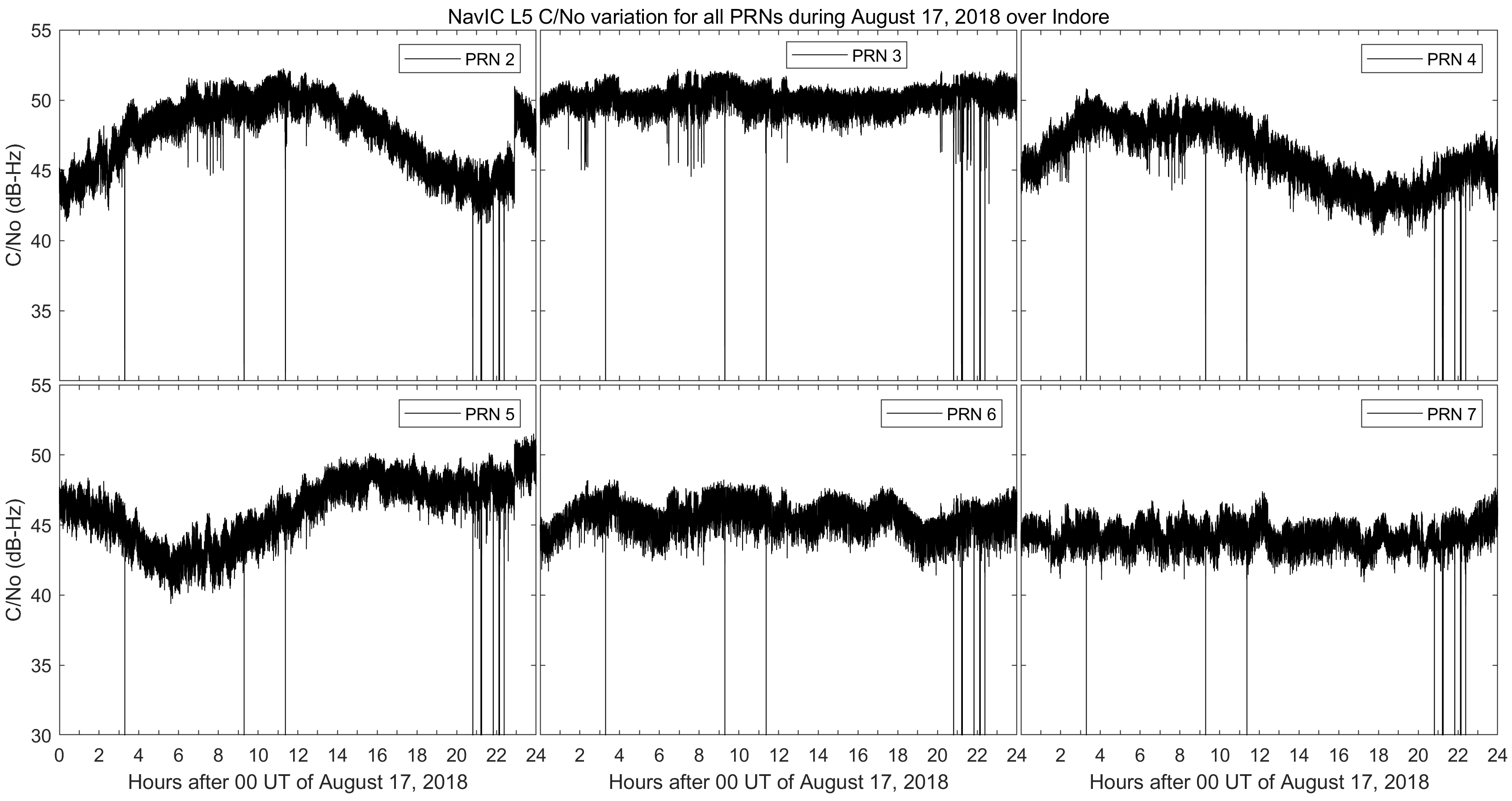}
\caption{The C/N$_o$ (dB-Hz) variation during the entire day of August 17, 2018, over Indore, as observed by the L5 signal of NavIC satellite PRNs 2-7.} 
\label{fig14}
\end{figure}

Very similar to the other two cases, Figure \ref{fig14} shows the C/N$_o$ variation (dB-Hz) for the entire day of August 17, 2018, as observed by the L5 signal of NavIC. Drops in the C/N$_o$ have been observed at multiple time stamps throughout the day by all the PRNs.

\begin{figure}[ht]
\centering
\noindent\includegraphics[width=4.5in,height=4.5in]{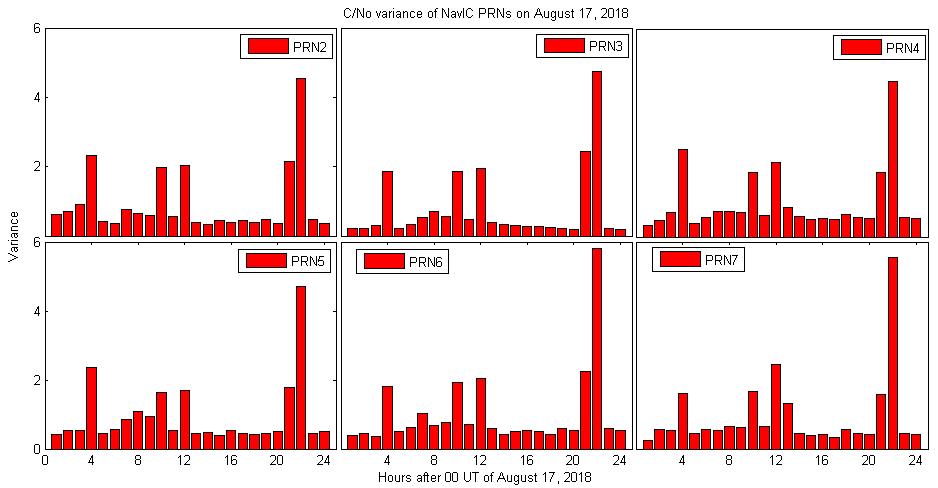}
\caption{The hourly binned variance plots of C/N$_o$ for all PRNs of NavIC on August 17, 2018, over Indore.} 
\label{fig15}
\end{figure}

However, for the verification of the significance of these C/N$_o$ drops, Figure \ref{fig15} shows the hourly binned variance plots of all the PRNs of NavIC during the entire day of August 17, 2018. The time bin of 21-22 UT for all the PRNs shows the most significant rise and hence maximum variation among all the bins of this day.

\begin{figure}[ht]
\centering
\noindent\includegraphics[width=3in,height=2in]{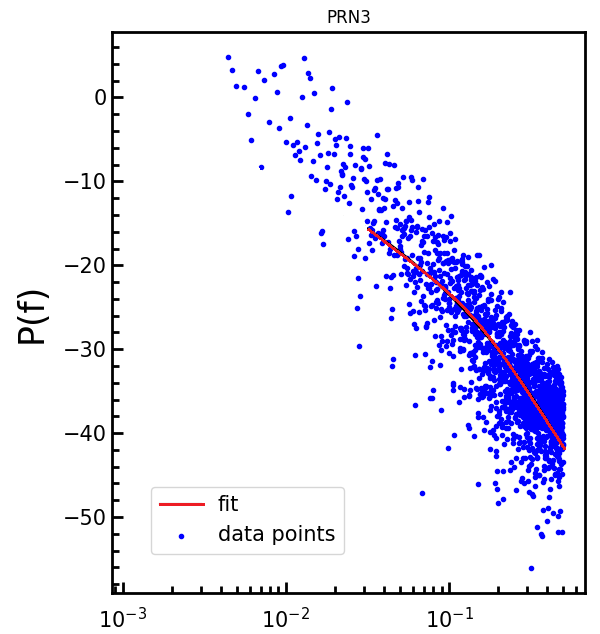}
\noindent\includegraphics[width=3in,height=2in]{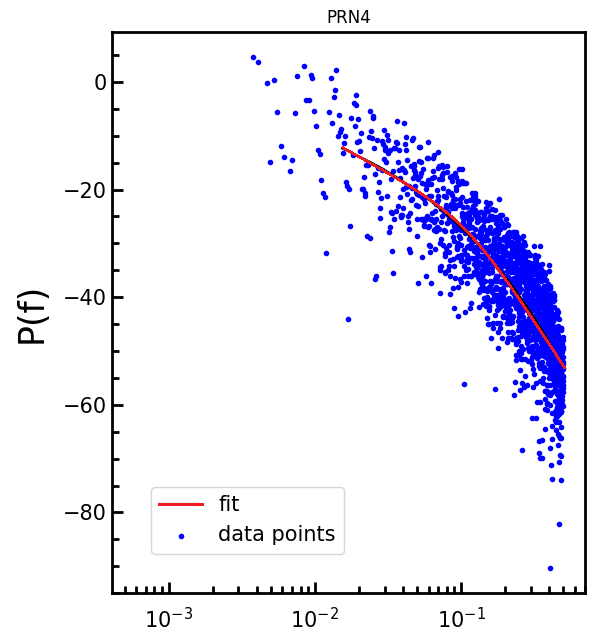}
\noindent\includegraphics[width=3in,height=2in]{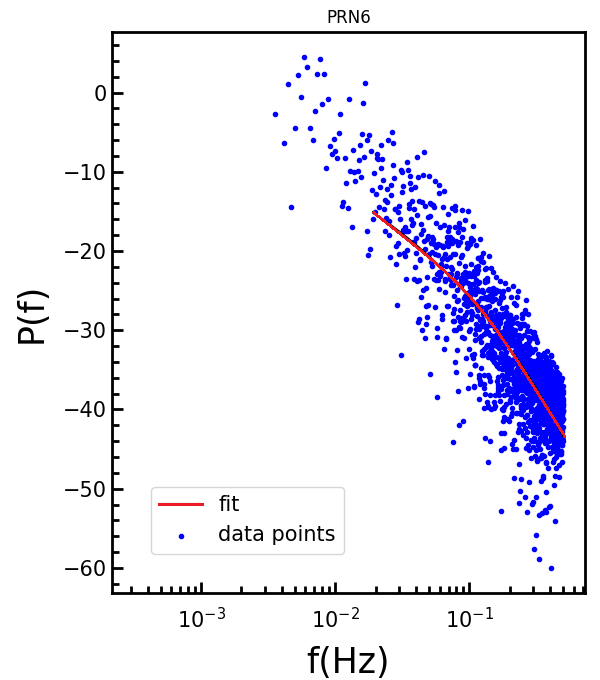}
\caption{The PSD variations with the least square fit (red solid line) correspond to the time bins of C/N$_o$ variation of Figure \ref{fig15}.}
\label{fig16}
\end{figure}

Figure \ref{fig16}, in a way similar to the previous cases, shows the PSD variation from all the NavIC satellites. Here, only PRNs 3, 4, and 6 show the power law variation and hence are taken into consideration. The values of p for PRNs 3, 4, and 6 are 2.306$\pm$0.009, 2.518$\pm$0.011, and 2.322$\pm$0.008 respectively. The corresponding outer irregularity scale size is 0.502$\pm$0.071 km.

\clearpage

\clearpage

Finally, To understand the overall nature of the ionospheric variabilities as observed through NavIC and the corresponding irregularity scale sizes, during the entire analysis period from simultaneous observations over locations near the northern crest of EIA (Indore) and in between the crest and magnetic equator (Hyderabad), Table \ref{tab} summarizes the estimated values of the irregularity scale sizes, for all the 27 days consisting of events under strong, weak-to-moderate, and quiet geomagnetic storm intensities, all of which had occurred in different seasons over one year. The corresponding solar radio flux values (F10.7 in s.f.u) have also been presented in the same table to show the solar activity level during these days. The range of the outer scale sizes observed implies the presence of larger (order of a few km) and smaller (order of a few 100 m) scale structures over both locations. The overall variations show that the scale sizes during the stronger event had been around the order of a few km whereas, for the weak-to-moderate, and quiet-time conditions, they were of the order of a few hundred meters.

\clearpage

\begin{table}[ht]
\caption{The outer scale sizes of irregularities for all the 27 days of observation when the L5 C/N$_o$ drops were observed under different geomagnetic conditions over the two locations: Indore (L$_{Indr}$) and Hyderabad (L$_{Hyde}$) are shown below. The 10.7 cm solar radio flux values (F10.7 in s.f.u) are indicated to show the corresponding solar activity level.}
\centering
\begin{center}
\resizebox{0.75\textwidth}{0.38\textheight}{
\begin{tabular}{|c|c|c|c|c|}
\hline
Cases      & F10.7 (s.f.u) & Storm Intensity & L$_{Indr}$ (km) & L$_{Hyde}$ (km) \\ \hline
08/09/2017 & 188.5         & strong          & 5.120$\pm$0.011 & 5.150$\pm$0.001 \\ \hline
16/09/2017 & 72.9          & moderate        & 0.507$\pm$0.001 & 0.504$\pm$0.086 \\ \hline
27/09/2017 & 91.3          & moderate        & 0.511$\pm$0.010 & 0.499$\pm$0.015 \\ \hline
12/10/2017 & 69.9          & moderate        & 0.517$\pm$0.032 & 0.514$\pm$0.021 \\ \hline
24/10/2017 & 76.7          & moderate        & 0.508$\pm$0.007 & 0.506$\pm$0.098 \\ \hline             
25/10/2017 & 77.9          & moderate        & 0.480$\pm$0.006 & 0.505$\pm$0.093 \\ \hline          
01/12/2017 & 68.3          & quiet           & 0.507$\pm$0.005 & 0.512$\pm$0.014 \\ \hline
04/12/2017 & 66.4          & weak            & 0.504$\pm$0.085 & 0.489$\pm$0.026 \\ \hline
14/12/2017 & 69.8          & quiet           & 0.506$\pm$0.007 & 0.519$\pm$0.039 \\ \hline
27/12/2017 & 68.6          & quiet           & 0.514$\pm$0.015 & 0.502$\pm$0.026 \\ \hline
16/02/2018 & 69.8          & quiet           & 0.524$\pm$0.020 & 0.519$\pm$0.021 \\ \hline
19/02/2018 & 67.5          & weak            & 0.521$\pm$0.001 & 0.502$\pm$0.056 \\ \hline
20/02/2018 & 66.3          & quiet           & 0.499$\pm$0.048 & 0.498$\pm$0.006 \\ \hline
21/02/2018 & 68.7          & quiet           & 0.511$\pm$0.085 & 0.500$\pm$0.018 \\ \hline
22/02/2018 & 67.0          & weak            & 0.514$\pm$0.007 & 0.489$\pm$0.085 \\ \hline
24/02/2018 & 66.8          & weak            & 0.490$\pm$0.006 & 0.489$\pm$0.026 \\ \hline
25/02/2018 & 65.9          & quiet           & 0.494$\pm$0.090 & 0.481$\pm$0.019 \\ \hline
27/02/2018 & 66.6          & moderate        & 0.512$\pm$0.000 & 0.489$\pm$0.099 \\ \hline
28/02/2018 & 67.5          & quiet           & 0.510$\pm$0.097 & 0.489$\pm$0.036 \\ \hline
01/03/2018 & 66.0          & quiet           & 0.495$\pm$0.091 & 0.517$\pm$0.030 \\ \hline
02/03/2018 & 66.6          & quiet           & 0.522$\pm$0.005 & 0.509$\pm$0.002 \\ \hline
13/03/2018 & 67.7          & quiet           & 0.507$\pm$0.095 & 0.499$\pm$0.015 \\ \hline
22/03/2018 & 68.0          & quiet           & 0.514$\pm$0.001 & 0.519$\pm$0.008 \\ \hline
28/04/2018 & 71.1          & quiet           & 0.486$\pm$0.011 & 0.508$\pm$0.054 \\ \hline
28/05/2018 & 78.9          & quiet           & 0.507$\pm$0.009 & 0.502$\pm$0.089 \\ \hline
05/06/2018 & 73.4          & quiet           & 0.517$\pm$0.005 & 0.504$\pm$0.001 \\ \hline
17/08/2018 & 69.0          & weak            & 0.502$\pm$0.071 & 0.509$\pm$0.092 \\ \hline
\end{tabular}}
\end{center}
\label{tab}
\end{table}
\clearpage

\section{Discussion}

The low-latitude region over the globe is severely affected by the ionospheric scintillations, which degrades the performance of the satellite-based communication and navigational systems on which society is heavily dependent. The intermediate scale size irregularities (about a few 100 meters to a few km), which are the major sources of scintillations on the L-band and VHF transionospheric radio signals and scatter the high-frequency radio waves that are propagating through the low-to-equatorial ionosphere, are formed as a result of the R-T instability growth on the bottom side of the equatorial F-layer during the post-sunset local time. This scattering, in turn, creates a spatial pattern in the form of variations in the amplitude and phase of these signals on the receiver plane. Due to the relative movement of these irregularities with respect to the path of the signals that are transmitted from GNSS, the spatial patterns of amplitude and phase variations move past the receiver resulting in rapid temporal fluctuations in the phase and the amplitude of the signal that is being received by these ground-based GNSS receivers. On any given night, these intermediate-scale size irregularities associated with the EPBs can cover a large portion of the ionosphere over a particular longitude sector in the low-latitude regions (\cite{sc:27} and references therein).

It is well known that the F-region ionospheric irregularities exhibit power law spectra.  For various conditions of the ionosphere, the power law dependence is universal and has been verified by various researchers (\cite{sc:43,sc:44,sc:45,sc:46,sc:47} and references therein) using multiple measurements. A power law variation would signify the existence of some non-linear processes. These cascade-type processes would allow energy (the wind shears or solar heating) inflow to get redistributed across the various scales in addition to the larger inhomogeneities (energy-containing inhomogeneities of the largest size called the outer scale) getting split (due to the bending and stretching under the inertial forces) into the less energetic and smaller inhomogeneities. This process continues till the smallest inhomogeneity scale (inner scale) is reached, where the viscous forces dominate the inertial forces and the dissipation of energy takes place \citep{sc:42}.

In the phase screen model \citep{sc:15,sc:16} given by Rino, the spectral slopes obtained from the PSD analysis are considered as a key parameter for the accurate determination of the level or intensity of scintillation \citep{sc:48}. Looking back at the case studies in the present investigation, we observe that for the September 08, 2017 event, the values of the spectral slopes were around -3.6. This is close to the value of -11/3. The ionospheric structures that were present during that particular day produced a power law spectral nature closely resembling the Kolmogorov (-5/3) spectrum, the slope (-11/3) being the characteristic feature of that turbulence. This type of spectral slope gets generated by the injection of energy (geomagnetic field disturbances, instabilities due to solar forcing, etc.) into the system at large scales and the non-linear transfer of energy from the larger structures to the smaller ones, via the evolution of turbulence through the non-linear interactions between the fluctuations in plasma. Here the -11/3 slope suggests the process of anisotropic cascading of turbulent energy. This energy eventually dissipates, at smaller scales, via the mechanism of wave-particle interactions and/or ion-neutral collisions (\cite{sc:49,sc:50,sc:51,sc:52,sc:53,sc:42} and references therein). For the cases of September 16, 2017, and August 17, 2018, the spectral slopes were around -2.8 and -2.5 respectively, which is close to the value of -8/3. This type of spectral slope is observed in regions where small-scale irregularities are dominant as a result of instabilities (strongly anisotropic and intermittent turbulence) at these scales and suggest rapid/steeper decay of energy compared to a Kolmogorov turbulent cascade, at these smaller scales.

\begin{figure}[ht]
\centering
\noindent\includegraphics[width=5in,height=5in]{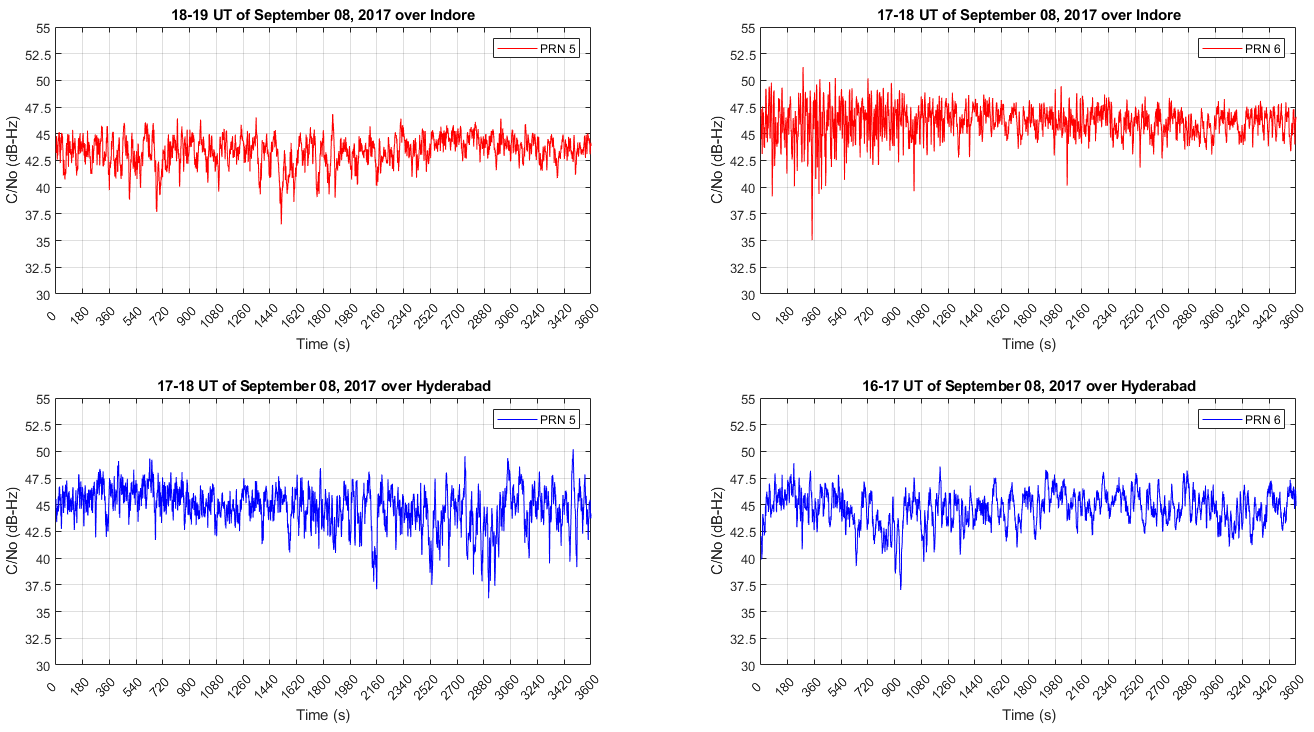}
\caption{The C/N$_o$ (dB-Hz) variation during significant time bins of September 08, 2017, as observed by the L5 signal of NavIC satellite PRNs 5 and 6 over both Indore and Hyderabad. For Indore, LT (h) = UT + 05:04 and for Hyderabad, LT (h) = UT + 05:14} 
\label{fig17}
\end{figure}

Furthermore, Figure \ref{fig17} shows the C/N$_o$ variations of NavIC PRNs 5 and 6 in red as observed over Indore and in blue as observed over Hyderabad for the respective bins when there are significant drops as observed from the variance plots (i.e. Figure \ref{fig4}). Now, as the F-region ionospheric irregularities over the geomagnetic equator get mapped along the geomagnetic field lines to latitudes away from the equator (near and around the crest regions), a time delay in scintillation occurrence, as a result of the movement of these irregularity structures, is expected when observing over a region (Indore) close to the crest in comparison to a region (Hyderabad) away from the crest. In Figure \ref{fig17}, the C/N$_o$ drops observed by PRNs 5 and 6 are at 18:25 UT (23:29 LT) and 17:06 UT (22:10 LT) respectively over Indore and the same at 17:49 UT (23:03 LT) and 16:16 UT (21:30 LT) over Hyderabad, suggesting a clear time delay, of scintillation occurrence and a poleward movement of the plasma irregularity structures, for about 36 minutes for PRN 5 and 50 minutes for PRN 6 over Indore with respect to that over Hyderabad.       

Next, the geographic locations of IPPs for PRNs 5 and 6 are (21.13$^\circ$N, 79.42$^\circ$E) and (21.58$^\circ$N, 71.28$^\circ$E) respectively over Indore, and (16.55$^\circ$N, 82.65$^\circ$E) and (16.26$^\circ$N, 74.45$^\circ$E) respectively over Hyderabad. Interestingly, on observing the time of C/N$_o$ drops over the two locations and the two PRNs in Figure \ref{fig17} along with the comparison with the longitudes of the PRN IPPs over these two locations, we can observe a westward propagation of the large-scale (km) plasma irregularity structures over these regions. As the IPPs of the two PRNs over Hyderabad are eastward in comparison to the same over Indore, and we observe scintillation over Hyderabad first, it is clear that the irregularity structures drifted westward. It is to be noted that the observed westward propagation of the irregularity structures in the present study is consistent with the study presented in \citep{sc:38} when an under-shielding condition of penetration electric field was present over these locations of observations and could have been a factor to have caused such westward drift of the irregularity structures.

\begin{figure}[ht]
\centering
\noindent\includegraphics[width=5in,height=5in]{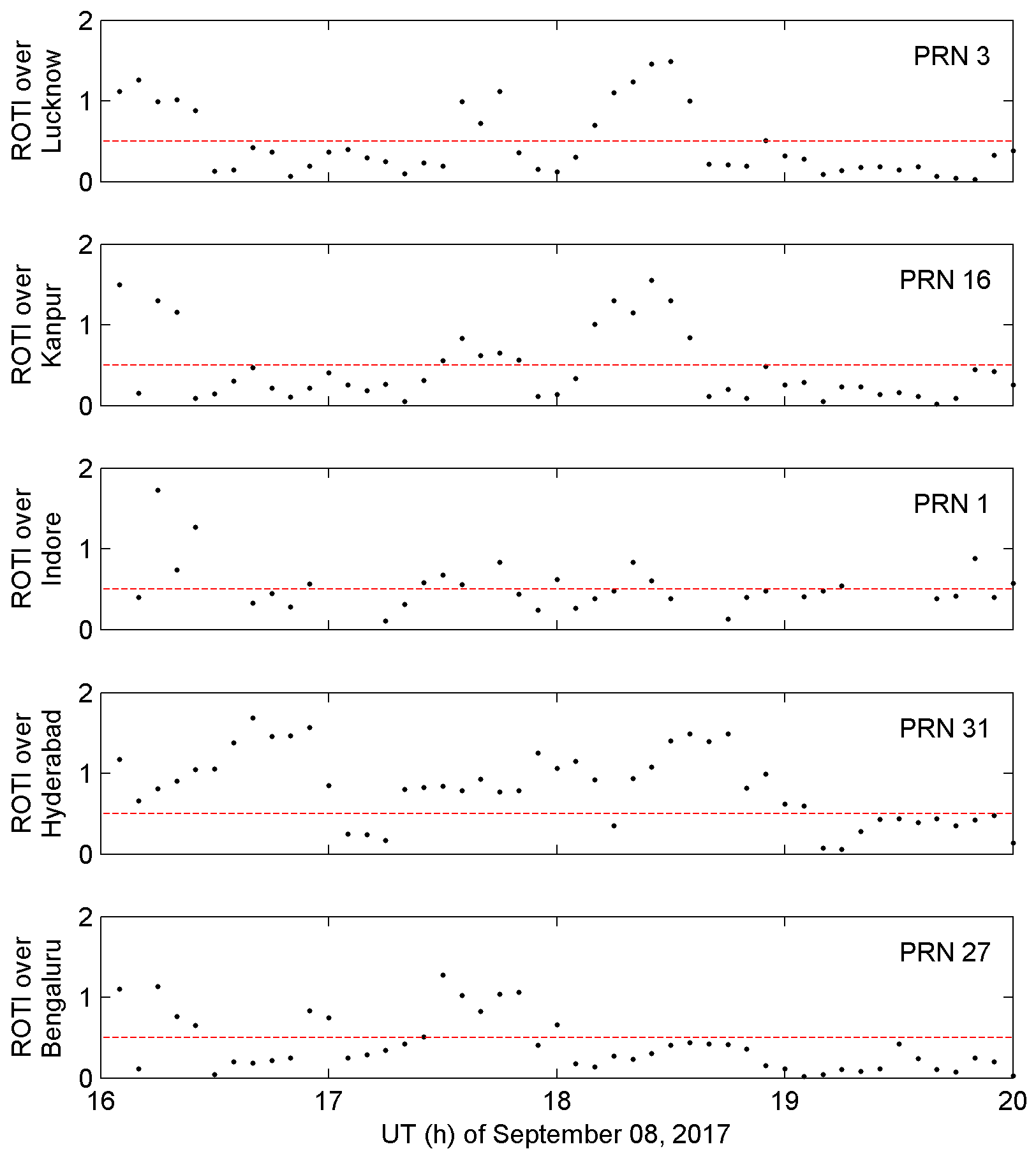}
\caption{The GNSS/GPS ROTI (TECU/min) variation along with the satellite PRNs during September 08, 2017. From top to bottom: The stations (Lucknow to Bengaluru) are arranged as one goes from beyond the EIA crest towards the dip equator. The values of 0.5 TECU/min have been indicated by red-dashed horizontal lines.} 
\label{fig18}
\end{figure}

An important proxy for scintillation occurrence and identification of the associated ionospheric disturbances and irregularity structures is the Rate of TEC Index (ROTI). This is the standard deviation of the Rate of change of TEC (ROT) over 5 minutes (\cite{sc:39,sc:40} and references therein). Further, if this ROTI shows values above 0.5 TECU/min, it would indicate the presence of a few km order of irregularity scale sizes \citep{sc:41}. Following these studies and to validate the presence of km scale sizes of irregularities on September 08, 2017, over the Indian sector, we show the variation of ROTI in Figure \ref{fig18}. The time duration is taken from 16:00 to 20:00 UT keeping in mind the occurrences of scintillations, shown earlier, between these periods on this day. These ROTI values are calculated from the GNSS/GPS data available at the IGS stations: Lucknow (26.91$^\circ$N, 80.96$^\circ$E, dip: 39.75$^\circ$N) and Kanpur (26.52$^\circ$N, 80.23$^\circ$E, dip: 38.52$^\circ$N) located beyond the EIA crest, and Hyderabad (17.42$^\circ$N, 78.55$^\circ$E, dip: 21.69$^\circ$N) and Bengaluru (13.02$^\circ$N, 77.57$^\circ$E, dip: 11.71$^\circ$N) located away from the crest and closest to dip equator respectively. The Indore (22.52$^\circ$N, 75.92$^\circ$E, dip: 32.23$^\circ$N and located near the EIA crest) data is obtained from the GNSS/GPS receiver available at IIT Indore. From this figure, there are multiple instances of ROTI values greater than 0.5 TECU/min, suggesting the presence of irregularities of the order of a few km that caused scintillations in the L-band signals of these satellites. It is to be noted that upon similar analysis (not shown) of the ROTI from these stations for the other events, we could observe the ROTI values to be well below 0.5 TECU/min, suggesting much smaller (few 100 m) irregularity scale sizes were present.

\clearpage

\section{Summary}

The ionosphere over the Indian longitude sector is highly dynamic and geosensitive due to the presence of the northern crest of EIA and the magnetic equator. A study of the ionospheric irregularities, that have drastic impacts on the performance of the global navigational satellite systems in and around these locations under geomagnetically active conditions, is essential. The detailed study of these ionospheric irregularities brings forward important aspects to understanding ionospheric physics and related processes. To address this issue, the present work investigated the simultaneous observations of low-latitude ionospheric irregularities over locations chosen to cover the zones of both the northern crest of EIA (Indore) and in between the crest and the magnetic equator (Hyderabad). This study utilized the L5 signal C/N$_o$ variations from a set of GEO and GSO satellites of NavIC for determining the irregularity spectral slope and the corresponding outer irregularity scale sizes, using the PSD analysis. The estimated scale sizes ranged from about 500 m over Indore and Hyderabad under weak-to-moderate and quiet-time conditions and about 5 km over both locations under the strong geomagnetic event. These scale sizes were further validated by the ROTI variations. Further, a time delay of scintillation occurrence over Indore, with values of 36 minutes and 50 minutes for NavIC PRNs 5 and 6 respectively, from scintillation occurrence at Hyderabad, was also observed. This observation suggested a poleward evolution of the irregularity structures on the event day. Furthermore, a westward propagation of these km-scale irregularity structures, was observed. This observation is consistent with the observations shown in the work by \citep{sc:38}. Therefore, this study brought forward the usefulness of monitoring the propagation and evolution of low-latitude ionospheric irregularities with simultaneous observations, from a region near the crest and another away from it, utilizing observations from a continuously available set of GEO and GSO navigation satellite systems.

\section*{Acknowledgments} 

SC acknowledges the Space Applications Centre (SAC), ISRO for providing a research fellowship under the project NGP-17 during his doctoral tenure. The authors acknowledge SAC/ISRO for providing the NavIC receivers to the Department of Astronomy, Astrophysics and Space Engineering (DAASE), IIT Indore under this project. AD acknowledges the use of facilities procured through the funding via the Department of Science and Technology (DST), Government of India (GoI) sponsored DST-FIST grant no. SR/FST/PSII/2021/162 (C) awarded to the DAASE, IIT Indore. The authors also acknowledge the Department of ECE, Osmania University for providing us with the NavIC data of Hyderabad. The authors also acknowledge Prof. Gopi K. Seemala of the Indian Institute of Geomagnetism (IIG), Navi Mumbai, India for providing the software available from https://seemala.blogspot.com/2017/09/gps-tec-program-ver-295.html?m=1 to analyze the IGS data available at http://sopac-csrc.ucsd.edu/index.php/data-download/. The authors acknowledge the valuable comments and suggestions from all the reviewers that have significantly enriched the manuscript's quality.  SC also acknowledges Aishrila Mazumder for fruitful discussions. Further acknowledgments go to the World Data Center for Geomagnetism, Kyoto at https://wdc.kugi.kyoto-u.ac.jp/kp/index.html for the K$_p$ index and the SPDF OMNIWeb database at https://omniweb.gsfc.nasa.gov/, for the high-resolution (1-minute) IMF B$_z$, V$_{sw}$, $\rho_{sw}$, and the SYM-H data.

\bibliographystyle{model2-names.bst}\biboptions{authoryear}
\bibliography{SC-P4.bib}

\end{document}